\documentclass[aps,prb,twocolumn,groupedaddress,citeautoscript]{revtex4}
\usepackage[dvipdfmx]{graphicx}
\usepackage{color}
\usepackage{bm}
\begin{document}

\title{Weyl points and Dirac lines protected by multiple screw rotations
}


\author{Akira Furusaki       
}


\affiliation{Condensed Matter Theory Laboratory, RIKEN, Wako, Saitama, 351-0198, Japan}
\affiliation{RIKEN Center for Emergent Matter Science, Wako, Saitama, 351-0198, Japan}


\date{\today}

\begin{abstract}
In three-dimensional noncentrosymmetric materials
two-fold screw rotation symmetry forces
electron's energy bands to have Weyl points at which
two bands touch.
This is illustrated for space groups
No.~19 ($P2_12_12_1$) and No.~198 ($P2_13$),
which have three orthogonal screw rotation axes.
In the case of space groups
No.~61 ($Pbca$) and No.~205 ($P$a-3)
that have extra inversion symmetry,
Weyl points are promoted to four-fold degenerate line nodes
in glide-invariant planes.
The three-fold rotation symmetry present in the space groups
No.~198 and No.~205
allows Weyl and Dirac points, respectively,
to appear along its rotation axes in the Brillouin zone
and generates four-fold and six-fold degeneracy at the
$\Gamma$ point and R point, respectively.
\keywords{Weyl points \and Dirac points \and Nodal lines \and nonsymmorphic symmetry}
\end{abstract}

\maketitle

\section{Introduction}
\label{intro}

Topological states of matter have attracted a lot of attention since the
discovery of topological insulators \cite{HasanKane,QiZhang}.
A focus of very active recent studies is topological
semimetals \cite{Murakami,WanTurner,Young2012} that have gapless excitations
in the bulk with linear energy dispersion
(Weyl or Dirac fermions).
For example, first-principles calculations \cite{WengReview} and subsequent
experiments have discovered Cd$_3$As$_2$ and Na$_3$Bi,
Dirac semimetals with four-fold degenerate band-touching points
(Dirac points) \cite{Wang2012,Wang2013},
and TaAs, a Weyl semimetal with two-fold degenerate band-touching
points (Weyl points) \cite{Lv2015,Xu2015}.

It has been known that
nonsymmorphic crystal symmetries, such as screw rotation and
glide mirror, enforce
energy bands to stick together at some high symmetry points
when spin-orbit coupling is negligible.
The band degeneracies that are stable in the presence of spin-orbit coupling
has been discussed only recently
\cite{Sid,Steinberg,YangNagaosa,YoungKane,Fang2015,Watanabe2016,Bradlyn2016,Chen2016,Wieder,ChenFangReview,YangBojesen,Titus2016,Wang2017}.
In this paper we study the band structure of materials whose crystalline
symmetry is governed by nonsymmorphic space groups (SGs) 19, 61, 198, and 205,
which have multiple screw rotation symmetries.
Our study is motivated by recent experiments and first-principle calculations
on cubic chiral materials NiSbS and PdBiSe (SG198)\cite{Kakihana2015}
and CoSe$_2$ (SG205)\cite{Teruya2016}, which revealed complex Fermi
surface structures and band touchings that are characteristic of
nonsymmorphic crystals with strong spin-orbit coupling.


\section{Space groups No.~19 and No.~198}
\label{sec: 198}

We first discuss energy band structures
of electron systems with strong spin-orbit coupling in crystals
of SG19 and SG198.
(A discussion on the band topology for the SG19 in the absence of
spin-orbit coupling can be found in Ref.~\onlinecite{Bouhon}.)
Throughout this paper we assume that electron systems are invariant
under time-reversal transformation $\Theta$, which is an antiunitary
operator satisfying $\Theta^2=-1$.
The SG19 and SG198 correspond to
orthorhombic and cubic crystals, respectively.
We set the lattice constants to be unity so that
both SGs can be treated on equal footing.

The SG19 and SG198 have two-fold screw rotations
about the $x$, $y$, and $z$ axes,
\begin{eqnarray}
\widetilde{C}_{2x}
&:&
(x,y,z)\rightarrow\textstyle{(x+\frac12,-y+\frac12,-z)},
\label{C_2x}
\\
\widetilde{C}_{2y}
&:&
(x,y,z)\rightarrow\textstyle{(-x,y+\frac12,-z+\frac12)},
\label{C_2y}
\\
\widetilde{C}_{2z}
&:&
(x,y,z)\rightarrow\textstyle{(-x+\frac12,-y,z+\frac12)}.
\label{C_2z}
\end{eqnarray}
In addition, the SG198 has three-fold rotation
about the (1,1,1) axis,
\begin{equation}
C_3:(x,y,z)\rightarrow(z,x,y),
\label{C_3}
\end{equation}
and its cousins generated by multiplying $C_3$ and $\widetilde{C}_{2\alpha}$.
In spin-orbit coupled systems, all these transformations
involve rotations in the spin space as well.
For example, the action of $\widetilde{C}_{2\alpha}$ in the spin space
is represented by $i\sigma_\alpha$,
where $\sigma_\alpha$ are Pauli matrices ($\alpha=x,y,z$).
The three screw rotations are not independent,
as they obey the relation
\begin{equation}
\widetilde{C}_{2y}\widetilde{C}_{2z}
=-T_{(-1,0,0)}\widetilde{C}_{2x}
=-e^{-ik_x}\widetilde{C}_{2x},
\label{C_2y * C_2z => C_2x}
\end{equation}
where $T_{(n_x,n_y,n_z)}$ is a translation operator
\begin{equation}
T_{(n_x,n_y,n_z)}: (x,y,z)\to(x+n_x,y+n_y,z+n_z).
\end{equation}
The second equality in Eq.\ (\ref{C_2y * C_2z => C_2x})
holds when operators act on Bloch states with wave number
$\bm{k}=(k_x,k_y,k_z)$,
and the minus signs in Eq.\ (\ref{C_2y * C_2z => C_2x}) are due to
the Pauli spin algebra $(i\sigma_y)(i\sigma_z)=-i\sigma_x$.
Furthermore, Eqs.\ (\ref{C_2x})--(\ref{C_2z}) imply
that the product $\widetilde{C}_{2x}\widetilde{C}_{2y}\widetilde{C}_{2z}$
transforms the coordinate $(x,y,z)$ to itself.
Taking operations in the spin space into account, we obtain
\begin{equation}
\widetilde{C}_{2x}\widetilde{C}_{2y}\widetilde{C}_{2z}
=i\sigma_xi\sigma_yi\sigma_z=1.
\label{C_2x C_2y C_2z}
\end{equation}
Next, comparing the following product operators,
\begin{eqnarray}
\widetilde{C}_{2x}\widetilde{C}_{2y}&:&
(x,y,z)
\rightarrow
\textstyle{(-x+\frac12,-y,z-\frac12)},
\label{C_2x C_2y}\\
\widetilde{C}_{2y}\widetilde{C}_{2x}&:&
(x,y,z)
\rightarrow
\textstyle{(-x-\frac12,-y+1,z+\frac12)},
\end{eqnarray}
we find that
\begin{eqnarray}
\widetilde{C}_{2x}\widetilde{C}_{2y}&=&
-T_{(1,-1,-1)}\widetilde{C}_{2y}\widetilde{C}_{2x}
\nonumber\\
&=&
-e^{-i(k_x-k_y+k_z)}\widetilde{C}_{2y}\widetilde{C}_{2x},
\label{C_2x vs C_2y}
\end{eqnarray}
where the minus signs come from the anticommutation relation
between $\sigma_x$ and $\sigma_y$.
Similarly, we find
\begin{eqnarray}
\widetilde{C}_{2y}\widetilde{C}_{2z}
&=&
-e^{-i(k_x+k_y-k_z)}\widetilde{C}_{2z}\widetilde{C}_{2y},
\label{C_2y vs C_2z}
\\
\widetilde{C}_{2z}\widetilde{C}_{2x}
&=&
-e^{-i(-k_x+k_y+k_z)}\widetilde{C}_{2x}\widetilde{C}_{2z}.
\label{C_2x vs C_2z}
\end{eqnarray}

\subsection{Space group No.~19}

\begin{figure}
\begin{center}
\begin{picture}(200,175)(0,5)
\put(78,39){0}
\put(80,50){\vector(-2,-1){78}}
\put(-8,8){$k_x$}
\put(18,11){$\pi$}
\put(80,50){\vector(1,0){122}}
\put(77,175){$k_z$}
\put(70,150){$\pi$}
\put(80,50){\vector(0,1){120}}
\put(200,39){$k_y$}
\put(177,41){$\pi$}
\put(70,50){$\Gamma$}
\put(10,21){X}
\put(8,120){U}
\put(82,154){Z}
\put(120,10){S}
\put(184,53){Y}
\put(184,154){T}
\put(117,125){R}
\thicklines
\put(20,20){\line(1,0){100}}
\put(20,20){\line(0,1){100}}
\put(120,20){\line(0,1){100}}
\put(20,120){\line(1,0){100}}
\put(20,120){\line(2,1){60}}
\put(120,20){\line(2,1){60}}
\put(120,120){\line(2,1){60}}
\put(180,50){\line(0,1){100}}
\put(80,150){\line(1,0){100}}
%
%
\color{red}
\put(36,28){\circle*{5}}
\put(80,127){\circle*{5}}
\put(157,50){\circle*{5}}
\put(80,50){\circle*{5}}
\color{blue}
\put(20,120){\circle*{5}}
\put(120,20){\circle*{5}}
\put(180,150){\circle*{5}}
\put(100,120){\circle*{5}}
\end{picture}
\end{center}
\caption{
1/8 Brillouin zone for SG19,
where $0\le k_\alpha\le\pi$ ($\alpha=x,y,z$).
The red dots represent Weyl points.
The blue dots represent four-fold degenerate points or double Weyl points.
}
\label{fig: BZ/8 19}
\end{figure}
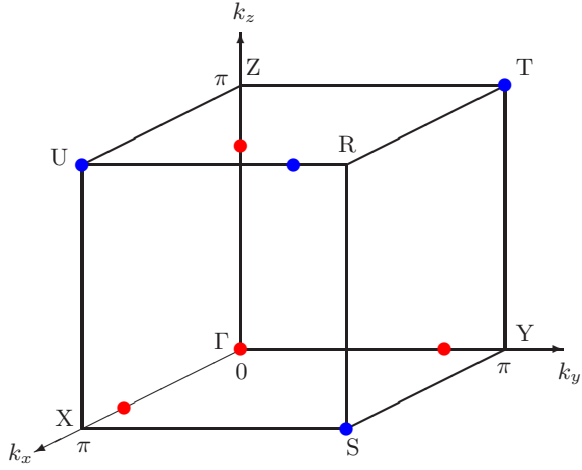

Let us first discuss energy band structures for SG19.
The invariant space of $\widetilde{C}_{2x}$,
a set of points in the Brillouin zone
that are invariant under the action of $\widetilde{C}_{2x}$,
is four lines parametrized as
\begin{eqnarray*}
\Gamma\mbox{-X}&:& (k_x,0,0),
\quad
\mbox{Y-S}: (k_x,\pi,0),\\
\mbox{Z-U}&:& (k_x,0,\pi),
\quad
\mbox{T-R}: (k_x,\pi,\pi),
\end{eqnarray*}
where $-\pi\le k_x\le\pi$;
see Fig.~\ref{fig: BZ/8 19}.
Bloch states with wave number $\bm{k}$ in the invariant space
are chosen to be eigenstates of both Hamiltonian and
$\widetilde{C}_{2x}$ operators.
The relation
$(\widetilde{C}_{2x})^2=T_{(1,0,0)}(i\sigma_x)^2=-e^{ik_x}$ implies that
the eigenvalues of $\widetilde{C}_{2x}$ are $\pm ie^{ik_x/2}$.
The Bloch states $|s,n,\bm{k}\rangle$ of $n$th energy band satisfy
\begin{equation}
\widetilde{C}_{2x}|\pm,n,\bm{k}\rangle
=\pm ie^{ik_x/2}
|\pm,n,\bm{k}\rangle.
\label{C_2x | >}
\end{equation}

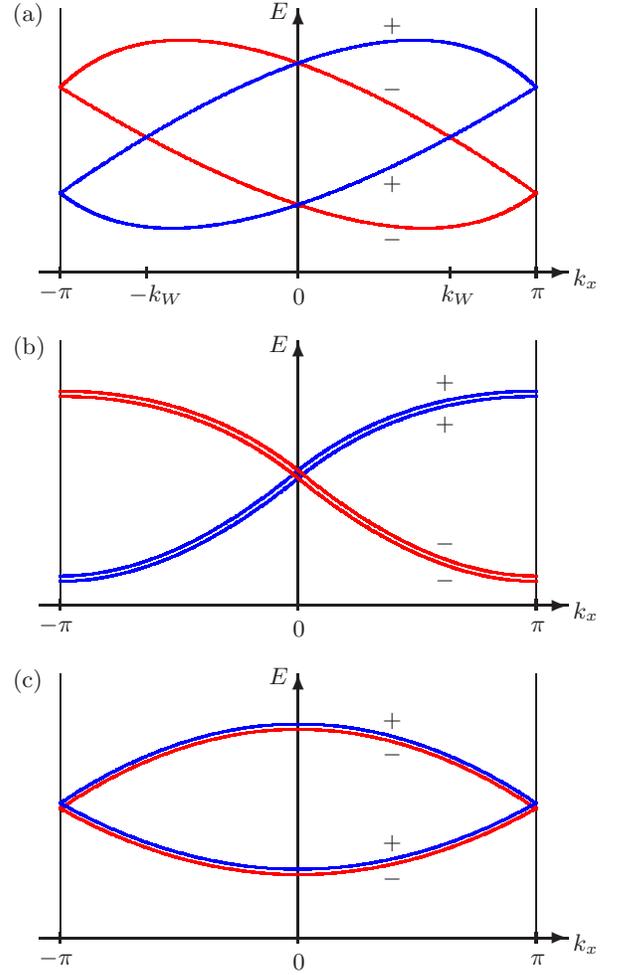
\begin{figure}
\begin{center}
\begin{picture}(250,125)(0,5)
\put(10,115){(a)}
\put(28,20){\line(0,1){100}}
\put(208,20){\line(0,1){100}}
\thicklines
\put(118,20){\vector(0,1){100}}
\put(107,116){$E$}
\put(20,20){\vector(1,0){200}}
\put(222,15){$k_x$}
\put(28,18){\line(0,1){4}}
\put(20,10){$-\pi$}
\put(118,18){\line(0,1){4}}
\put(116,8){$0$}
\put(208,18){\line(0,1){4}}
\put(206,10){$\pi$}
%
%
\put(150,111){$+$}
\put(150,87){$-$}
\put(150,51){$+$}
\put(150,30){$-$}
\put(175.5,18){\line(0,1){4}}
\put(172,9){$k_W$}
\put(60.5,18){\line(0,1){4}}
\put(54,9){$-k_W$}
\color{red}\qbezier(28,90)(78,140)(208,50)
\qbezier(28,90)(158,10)(208,50)
\color{blue}\qbezier(28,50)(78,10)(208,90)
\qbezier(28,50)(158,140)(208,90)
\end{picture}
%
\begin{picture}(250,125)(0,5)
\put(10,115){(b)}
\put(28,20){\line(0,1){100}}
\put(208,20){\line(0,1){100}}
\thicklines
\put(118,20){\vector(0,1){100}}
\put(107,116){$E$}
\put(20,20){\vector(1,0){200}}
\put(222,15){$k_x$}
\put(28,18){\line(0,1){4}}
\put(20,10){$-\pi$}
\put(118,18){\line(0,1){4}}
\put(116,8){$0$}
\put(208,18){\line(0,1){4}}
\put(206,10){$\pi$}
\put(170,102){$+$}
\put(170,86){$+$}
\put(170,41){$-$}
\put(170,27){$-$}
%
%
\color{blue}{\qbezier(208,99)(156,99)(118,68)}
\qbezier(208,101)(156,101)(118,71)
\qbezier(28,29)(70,29)(118,68)
\qbezier(28,31)(70,31)(118,71)
\color{red}{\qbezier(208,29)(166,29)(118,68)}
\qbezier(208,31)(166,31)(118,71)
\qbezier(28,99)(80,99)(118,68)
\qbezier(28,101)(80,101)(118,71)
\end{picture}
%
\begin{picture}(250,125)(0,5)
\put(10,115){(c)}
\put(28,20){\line(0,1){100}}
\put(208,20){\line(0,1){100}}
\thicklines
\put(118,20){\vector(0,1){100}}
\put(107,116){$E$}
\put(20,20){\vector(1,0){200}}
\put(222,15){$k_x$}
\put(28,18){\line(0,1){4}}
\put(20,10){$-\pi$}
\put(118,18){\line(0,1){4}}
\put(116,8){$0$}
\put(208,18){\line(0,1){4}}
\put(206,10){$\pi$}
\put(150,100){$+$}
\put(150,87){$-$}
\put(150,53.5){$+$}
\put(150,40){$-$}
%
\color{red}{\qbezier(28,69)(118,129)(208,69)}
\qbezier(28,69)(118,19)(208,69)
\color{blue}{\qbezier(28,71)(118,131)(208,71)}
\qbezier(28,71)(118,21)(208,71)
%
%
\end{picture}
\end{center}
\caption{
Schematic band structures for SG19.
The energy bands with ``$+$'' (blue) have the $\widetilde{C}_{2x}$-eigenvalue
$+ie^{ik_x/2}$,
and those with ``$-$'' (red) have the $\widetilde{C}_{2x}$-eigenvalue
$-ie^{ik_x/2}$.
(a) Along the invariant line $(k_x,0,0)$.
Weyl points exist at $k_x=0,\pm k_W$.
(b) Along the invariant line $(k_x,\pi,\pi)$.
(c) Along the invariant line $(k_x,\pi,0)$ and $(k_x,0,\pi)$.
In (b) and (c) the upper and lower bands are two-fold degenerate.
}
\label{fig: basic bands}
\end{figure}

The time-reversal operator $\Theta$ is an antiunitary operator
and commutes with any operator of space group transformations.
Multiplying $\Theta$
on both sides of Eq.\ (\ref{C_2x | >}) yields
\begin{equation}
\widetilde{C}_{2x}\Theta|\pm,n,\bm{k}\rangle
=\mp ie^{-ik_x/2}\Theta|\pm,n,\bm{k}\rangle.
\label{C_2x Theta}
\end{equation}
We see that the $\widetilde{C}_{2x}$
eigenvalues of a Kramers pair,
$|s,n,\bm{k}\rangle$ and $\Theta|s,n,\bm{k}\rangle$,
are different ($+i$ and $-i$) at $k_x=0$
and equal ($+1$ or $-1$) at $k_x=\pm\pi$.
This observation leads to the energy band structure
shown schematically in Fig.~\ref{fig: basic bands}(a).
Note that the bands with the $\widetilde{C}_{2x}$ eigenvalue
$+ie^{ik_x/2}$ (blue)
are smoothly connected at $k_x=\pi$ to the bands with
$\widetilde{C}_{2x}=-ie^{ik_x/2}$ (red) at $k_x=-\pi$,
and vice versa.
The bands with different $\widetilde{C}_{2x}$ eigenvalues
can cross and form Weyl points at $k_x=\pm k_W$ and at $k_x=0$.
We note that the schematic band structure shown
in Fig.~\ref{fig: basic bands}(a)
is the simplest one with a minimal number of band crossings/touchings.
The band structure of real materials can be more complicated with deformed
band dispersion (while keeping the symmetry) and have
more Weyl points that are generated by crossing energy bands
with different colors ($\widetilde{C}_{2x}$ eigenvalues).
The total monopole charge of Weyl points is unchanged by
such deformation of band structures.
Such band structures as shown in Fig.~\ref{fig: basic bands}(a)
should be realized along the invariant line $\Gamma$-X,
and also along the $\Gamma$-Y and $\Gamma$-Z lines;
for the latter two lines the role of $\widetilde{C}_{2x}$ is
played by $\widetilde{C}_{2y}$ and $\widetilde{C}_{2z}$,
respectively.
However, the band structure along the Y-S, Z-U, and T-R lines
are different from Fig.~\ref{fig: basic bands}(a),
as we discuss below.

The combination of the time-reversal and screw rotation transformations
guarantees Kramers degeneracy on the boundaries of the Brillouin zone,
$\bm{k}=(\pm\pi,k_y,k_z)$, $(k_x,\pm\pi,k_z)$, or $(k_x,k_y,\pm\pi)$,
where $-\pi\le k_\alpha\le\pi$.
For example, any $\bm{k}$ point on the $k_x=\pi$ plane is
invariant under the product operation
$\Theta\widetilde{C}_{2x}$ satisfying
\begin{equation}
(\Theta\widetilde{C}_{2x})^2=\Theta^2(\widetilde{C}_{2x})^2
=e^{ik_x}=-1.
\end{equation}
The presence of the antiunitary operator
$\Theta\widetilde{C}_{2x}$ squaring to $-1$ implies that
any energy level on the $k_x=\pi$ plane must be Kramers degenerate.
Similar arguments hold for the $k_y=\pi$ plane and the $k_z=\pi$ plane.
Hence the energy bands along Y-S, Z-U, and T-R lines
must be at least doubly degenerate.
On the contrary, Bloch states are generically non-degenerate
on the $\Gamma$-X line,
where $(\Theta\widetilde{C}_{2y})^2=(\Theta\widetilde{C}_{2z})^2=+1$.

Using Eqs.\ (\ref{C_2x vs C_2y}) and (\ref{C_2x | >}),
we obtain
\begin{equation}
\widetilde{C}_{2x}\widetilde{C}_{2y}|\pm,n,\bm{k}\rangle
=
\mp ie^{-\frac{i}{2}k_x+i(k_y-k_z)}\widetilde{C}_{2y}|\pm,n,\bm{k}\rangle,
\label{C_2y|phi>}
\end{equation}
where $\bm{k}$ is on a $\widetilde{C}_{2x}$-invariant line.
For $\bm{k}\in\Gamma$-X line,
$\widetilde{C}_{2y}|\pm,n,\bm{k}\rangle$
have the eigenvalues $\mp ie^{-ik_x/2}$,
in agreement with
the band structure of Fig.~\ref{fig: basic bands}(a)
(note that $\widetilde{C}_{2y}$ flips the sign of $k_x$).
Applying $\Theta$ to Eq.\ (\ref{C_2y|phi>}) yields
\begin{equation}
\widetilde{C}_{2x}\Theta\widetilde{C}_{2y}|\pm, n,\bm{k}\rangle
\!=\!
\pm ie^{\frac{i}{2}k_x-i(k_y-k_z)}
\Theta\widetilde{C}_{2y}|\pm, n,\bm{k}\rangle,
\label{ThetaC_2y|phi>}
\end{equation}
implying that a Kramers pair of states
$|s,n,\bm{k}\rangle$ and $\Theta\widetilde{C}_{2y}|s,n,\bm{k}\rangle$
have $\widetilde{C}_{2x}$-eigenvalues
$ise^{\frac{i}{2}k_x}$ and
\linebreak
$ise^{\frac{i}{2}k_x-i(k_y-k_z)}$,
respectively ($s=\pm$).
This leads to the band structures of Fig.~\ref{fig: basic bands}(b)
for the Bloch states on the T-R line.
By contrast, the energy bands along
the Y-S and Z-U lines
should have the structure shown schematically in Fig.~\ref{fig: basic bands}(c).
Both upper and lower bands are two-fold degenerate
in Figs.~\ref{fig: basic bands}(b) and \ref{fig: basic bands}(c).
The four-fold degenerate band crossing occurs at $k_x=0$ (T)
in Fig.~\ref{fig: basic bands}(b) and at $k_x=\pm\pi$ (S, U)
in Fig.~\ref{fig: basic bands}(c).

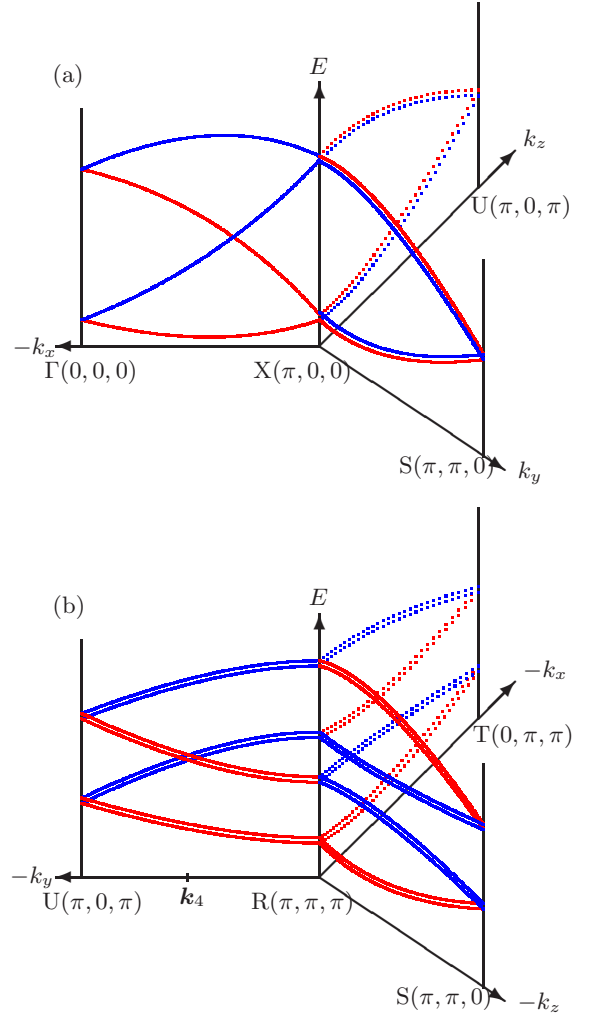
\begin{figure}
\begin{center}
%
%
\begin{picture}(250,185)(-10,-33)
\put(17,120){(a)}
\thicklines
\put(28,20){\line(0,1){90}}
\put(14,8){$\Gamma(0,0,0)$}
\put(178,80){\line(0,1){70}}
\put(175,70.5){U$(\pi,0,\pi)$}
\put(180,-22){\line(0,1){75}}
\put(148,-28){S$(\pi,\pi,0)$}
\put(118,20){\vector(0,1){100}}
\put(114,123){$E$}
\put(93,8){X$(\pi,0,0)$}
\put(118,20){\vector(-1,0){100}}
\put(2,17){$-k_x$}
\put(118,20){\vector(1,1){74}}
\put(195,96){$k_z$}
\put(118,20){\vector(3,-2){70}}
\put(193,-30){$k_y$}
\color{red}
\qbezier(28,87)(80,75)(118,32)
\qbezier(28,30)(80,17)(118,30)
\color{blue}
\qbezier(28,87)(80,110)(118,92)
\qbezier(28,30)(80,50)(118,90.5)
\color{red}
\qbezier[30](118,92)(140,117)(178,117)
\qbezier[40](118,32)(138,47)(178,117)
\color{blue}
\qbezier[30](118,90)(140,115)(178,115)
\qbezier[40](118,30)(140,45)(178,114.5)
\color{red}
\qbezier(118,92)(142,82)(180,17)
\qbezier(118,30)(140,10)(180,15)
\color{blue}
\qbezier(118,90)(140,80)(180,15)
\qbezier(118,33)(140,12)(180,17)
\end{picture}
%
%
\begin{picture}(250,200)(-10,-33)
\put(17,120){(b)}
\thicklines
\put(28,20){\line(0,1){90}}
\put(13,8){U$(\pi,0,\pi)$}
\put(178,80){\line(0,1){80}}
\put(176,71){T$(0,\pi,\pi)$}
\put(180,-22){\line(0,1){85}}
\put(147,-28){S$(\pi,\pi,0)$}
\put(118,20){\vector(0,1){100}}
\put(114,123){$E$}
\put(92,8){R$(\pi,\pi,\pi)$}
\put(118,20){\vector(-1,0){100}}
\put(1,18){$-k_y$}
\put(118,20){\vector(1,1){74}}
\put(195,96){$-k_x$}
\put(118,20){\vector(3,-2){70}}
\put(193,-30){$-k_z$}
\put(68,18.5){\line(0,1){3}}
\put(65,10){$\bm{k}_4$}
\color{blue}
\qbezier(28,82)(80,102)(118,102)
\qbezier(28,80)(80,100)(118,100)
\qbezier(28,50)(80,75)(118,75)
\qbezier(28,48)(80,73)(118,73)
\color{red}
\qbezier(28,82)(80,58)(118,58)
\qbezier(28,80)(80,56)(118,56)
\qbezier(28,50)(80,35)(118,35)
\qbezier(28,48)(80,33)(118,33)
%
\color{red}
\qbezier[30](118,75)(140,84)(178,130)
\qbezier[30](118,73)(140,82)(178,128)
\qbezier[40](118,35)(138,47)(178,100)
\qbezier[40](118,33)(140,45)(178,98)
\color{blue}
\qbezier[30](118,102)(140,120)(178,130)
\qbezier[30](118,100)(140,118)(178,128)
\qbezier[35](118,58)(138,75)(178,100)
\qbezier[35](118,56)(138,73)(178,98)
%
\color{red}
\qbezier(118,102)(142,96)(180,40)
\qbezier(118,100)(141,94)(180,38)
\qbezier(118,35)(140,12)(180,10)
\qbezier(118,33)(140,10)(180,8)
\color{blue}
\qbezier(118,75)(140,58)(180,40)
\qbezier(118,73)(140,56)(180,38)
\qbezier(118,58)(140,50)(180,10)
\qbezier(118,56)(139,48)(180,8)
\end{picture}
\end{center}
\caption{
Band structures for SG19
(a) along the $\Gamma$-X, X-S, and X-U lines
and (b) along the R-S, R-T, and R-U lines.
The energy bands with blue (red) color have the $\widetilde{C}_{2\alpha}$
eigenvalue $+ie^{ik_\alpha/2}$ ($-ie^{ik_\alpha/2}$),
where $\alpha=x$, $y$, or $z$.
}
\label{fig: connection}
\end{figure}

Next we discuss how energy bands are connected
at high symmetry points X and R.
Equations (\ref{C_2x vs C_2y})--(\ref{C_2y vs C_2z}) imply that
\begin{equation}
\{\widetilde{C}_{2x},\widetilde{C}_{2y}\}
=\{\widetilde{C}_{2y},\widetilde{C}_{2z}\}
=\{\widetilde{C}_{2z},\widetilde{C}_{2x}\}
=0
\end{equation}
at the time-reversal invariant momenta
with $k_x+k_y+k_z=0$ (mod $2\pi$),
while
\begin{equation}
[\widetilde{C}_{2x},\widetilde{C}_{2y}]
=[\widetilde{C}_{2y},\widetilde{C}_{2z}]
=[\widetilde{C}_{2z},\widetilde{C}_{2x}]
=0
\end{equation}
at the time-reversal invariant momenta
with $k_x+k_y+k_z=\pi$ (mod $2\pi$).
Thus, Bloch states at the X, Y, Z, and R points
can be chosen to be simultaneous eigenstates
of $\widetilde{C}_{2x}$, $\widetilde{C}_{2y}$, and $\widetilde{C}_{2z}$,
under the condition (\ref{C_2x C_2y C_2z}).
For example, 
doubly degenerate Bloch states at the X point $(\pi,0,0)$
have the eigenvalues
$(\widetilde{C}_{2x},\widetilde{C}_{2y},\widetilde{C}_{2z})
=(+1,\pm i,\mp i)$ or
$(-1,\pm i,\pm i)$.
The energy bands along the $\Gamma$-X line (Fig.~\ref{fig: basic bands}(a)) 
and those along the X-S, and X-U lines (Fig.~\ref{fig: basic bands}(c))
are connected at the X point as schematically shown
in Fig.~\ref{fig: connection}(a).

The two-fold degenerate energy bands along the R-S, R-T, and R-U lines
with the dispersion of the type shown in Fig.~\ref{fig: basic bands}(b)
are connected at the R point $(\pi,\pi,\pi)$,
where the possible combinations of the eigenvalues are
$(\widetilde{C}_{2x},\widetilde{C}_{2y},\widetilde{C}_{2z})=(1,-1,-1)$,
$(-1,1,-1)$, $(-1,-1,1)$, and $(1,1,1)$.
Consistent connection of bands requires at least 8 bands \cite{Watanabe2016},
as shown schematically
in Fig.~\ref{fig: connection}(b).
We note that a four-fold degenerate band crossing [at
$\bm{k}=\bm{k}_4$ in Fig.~\ref{fig: connection}(b)] always occurs
on at least one of the R-S, R-T, and R-U lines.
When $(8n+4)$ bands are filled, a four-fold degenerate band crossing point
should be located at the Fermi energy,
provided that there are no electron and hole pockets.


\subsection{Space group No.~198}

The SG198 has the $C_3$ symmetry defined in Eq.\ (\ref{C_3})
as a generator in addition to those of the SG19.
As a result the X, Y, and Z points become equivalent and called X,
while the S, T, and U points are renamed M.
Furthermore, the $C_3$ symmetry gives rise to
band crossings on the $\Gamma$-R line.

\begin{figure}
\begin{center}
\begin{picture}(250,200)(-10,-33)
\thicklines
\put(28,20){\line(0,1){90}}
\put(13,8){M$(\pi,0,\pi)$}
\put(178,80){\line(0,1){80}}
\put(174,70){M$(0,\pi,\pi)$}
\put(180,-22){\line(0,1){85}}
\put(145,-28.5){M$(\pi,\pi,0)$}
\put(118,20){\vector(0,1){100}}
\put(114,123){$E$}
\put(92,7.5){R$(\pi,\pi,\pi)$}
\put(118,20){\vector(-1,0){100}}
\put(1,18){$-k_y$}
\put(118,20){\vector(1,1){74}}
\put(195,96){$-k_x$}
\put(118,20){\vector(3,-2){70}}
\put(193,-30){$-k_z$}
\color{blue}
\qbezier(28,90)(80,102)(118,82)
\qbezier(28,88)(80,100)(118,80)
\qbezier(28,50)(80,75)(118,82)
\qbezier(28,48)(80,73)(118,80)
\color{red}
\qbezier(28,90)(80,80)(118,82)
\qbezier(28,88)(80,78)(118,80)
\qbezier(28,50)(80,35)(118,35)
\qbezier(28,48)(80,33)(118,33)
\color{red}
\qbezier[30](118,82)(140,84)(178,130)
\qbezier[30](118,80)(140,82)(178,128)
\qbezier[40](118,35)(138,47)(178,100)
\qbezier[40](118,33)(140,45)(178,98)
\color{blue}
\qbezier[30](118,82)(140,120)(178,130)
\qbezier[30](118,80)(140,118)(178,128)
\qbezier[35](118,82)(138,88)(178,100)
\qbezier[35](118,80)(138,86)(178,98)
\color{red}
\qbezier(118,82)(140,58)(180,40)
\qbezier(118,80)(140,56)(180,38)
\qbezier(118,35)(140,12)(180,10)
\qbezier(118,33)(140,10)(180,8)
\color{blue}
\qbezier(118,82)(142,78)(180,40)
\qbezier(118,80)(141,76)(180,38)
\qbezier(118,82)(140,50)(180,10)
\qbezier(118,80)(139,48)(180,8)
\end{picture}
\end{center}
\caption{
Band dispersion around the R point for SG198
and SG205.
The energy bands with blue (red) color have the $\widetilde{C}_{2\alpha}$
eigenvalue $+ie^{ik_\alpha/2}$ ($-ie^{ik_\alpha/2}$),
where $\alpha=x$, $y$, or $z$.
The $C_3$ symmetry about the R point is evident.
}
\label{fig: connection at R with C_3}
\end{figure}
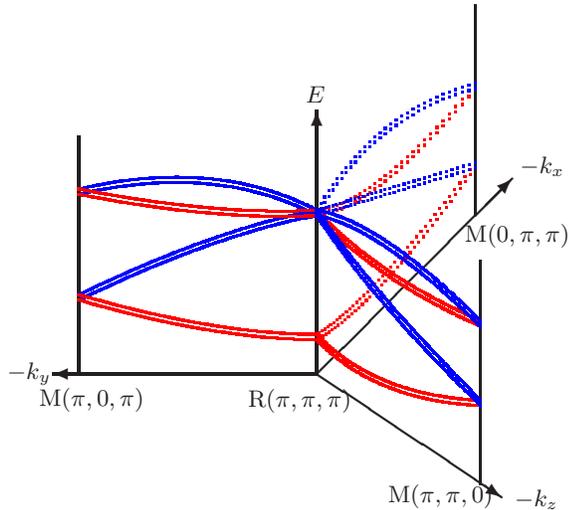

The algebraic relations among $C_3$ and $C_{2\alpha}$ symmetries,
\[
C_3\widetilde{C}_{2x}=\widetilde{C}_{2y}C_3,
\quad
C_3\widetilde{C}_{2y}=\widetilde{C}_{2z}C_3,
\quad
C_3\widetilde{C}_{2z}=\widetilde{C}_{2x}C_3,
\]
require that the six bands that have the eigenvalues
$(\widetilde{C}_{2x},\widetilde{C}_{2y},\widetilde{C}_{2z})
=(1,-1,-1),(-1,1,-1),(-1,-1,1)$
\linebreak
should be degenerate at the R point.
Therefore,
the energy levels at the R point are either
six-fold or two-fold degenerate \cite{Watanabe2016,Bradlyn2016}.
An example of the resulting band structure is shown
in Fig.~\ref{fig: connection at R with C_3}.
The $\bm{k}\cdot\bm{p}$ Hamiltonian at the six-fold degenerate
R point is discussed in Ref.~\onlinecite{Bradlyn2016}.

At the $\Gamma$ point the energy levels are either two-fold or four-fold
degenerate.
This can be understood as follows.
The symmetry operators can be represented as
$\widetilde{C}_{2\alpha}=\exp(i\pi S_\alpha)$,
$C_3=\exp[i(2\pi/3)(S_x+S_y+S_z)/\sqrt{3}]$, and
$\Theta=\exp(i\pi S_y)\mathcal{K}$, where
$\mathcal{K}$ is complex conjugation, and
$\bm{S}=(S_x,S_y,S_z)$ is an SU(2) spin operator in the
$S=\frac12$ and $S=\frac32$ irreducible representations
for two-fold and four-fold degenerate levels,
respectively.
The $C_3$ rotation symmetry is required in order for
the four-dimensional ($S=\frac32$) representation to be irreducible.

The low-energy $\bm{k}\cdot\bm{p}$ Hamiltonian at the $\Gamma$ point
in the four-dimensional representation ($S=\frac32$) is given by
\begin{eqnarray}
H_\Gamma&=&
a(k_xS_x+k_yS_y+k_zS_z)
+b(k_xS_x^3+k_yS_y^3+k_zS_z^3)
\nonumber\\
&&
+c[k_x\{S_x,S_y^2-S_z^2\}+k_y\{S_y,S_z^2-S_x^2\}
\nonumber\\
&&\qquad
+k_z\{S_z,S_x^2-S_y^2\}]
\end{eqnarray}
with three real parameters $a$, $b$, and $c$.
For a particular set of the parameters ($a=-\frac73 v$, $b=\frac43 v$, $c=0$),
it takes a simple form
\begin{eqnarray}
H_\Gamma^{(1)}&=&iv(k_x\widetilde{C}_{2x}
+k_y\widetilde{C}_{2y}+k_z\widetilde{C}_{2z})
\nonumber\\
&\equiv&v(k_x\sigma_x\tau_x-k_y\sigma_y\tau_x+k_z\sigma_z),
\end{eqnarray}
where $v$ is the velocity,
and $\sigma_\alpha$ and $\tau_\alpha$
are two sets of Pauli matrices.
In this case energy bands are doubly degenerate ($\tau_x=\pm1$) and
characterized by the Chern number $+1$ or $-1$;
this is a double Weyl point with the monopole charge 2.
For a different set of the parameters ($a=v$, $b=c=0$),
the $\bm{k}\cdot\bm{p}$ Hamiltonian
takes another simple form
\begin{equation}
H_\Gamma^{(2)}=v\bm{k}\cdot\bm{S}
\end{equation}
with $\bm{S}$ being the $S=\frac32$
spin operator.
The properties of this type of Hamiltonian are studied
in Ref.~\cite{Bradlyn2016} (for space groups other than SG198);
the four bands that are not degenerate
away from the $\Gamma$ point have the Chern numbers $\pm3$, $\pm1$,
and the total monopole charge is four \cite{Bradlyn2016}.
Hence, somewhere between these two Hamiltonians in the parameter space,
there has to be a topological phase transition where Chern numbers change.
Such topological phase transitions occur when $4a+13b=0$ or
when $16a^2+40ab+9b^2+64c^2=0$, i.e., when $H_\Gamma$ has two zeromodes
along the $\Gamma$-R or $\Gamma$-$Z$ direction.
The projection of the four-fold degenerate $\Gamma$ point on the
surface Brillouin zone emanates multiple Fermi arcs
corresponding to the total Chern number of filled bands.

Along the $\Gamma$-R line, the energy bands are classified in terms of
the $C_3$ eigenvalues: $-1$, $e^{i\pi/3}$, and $e^{-i\pi/3}$.
At the R point, sextuplets have
the three distinct $C_3$
eigenvalues twice each,
while doublets have the $C_3$ eigenvalues $(e^{i\pi/3},e^{-i\pi/3})$
or $(-1,-1)$.
At the $\Gamma$ point, the $C_3$ eigenvalues of
doublets are $(e^{i\pi/3},e^{-i\pi/3})$, and
those of quartets are
$(-1,e^{i\pi/3},e^{-i\pi/3},-1)$.
The crossing of bands with different $C_3$ eigenvalues gives rise to
Weyl points along the $\Gamma$-R line.

The crystal structure of NiSbS and PdBiSe has the symmetry of SG198,
and their band structures are shown in Figs.~8(a) and (c) in
Ref.~\onlinecite{Kakihana2015},
which were obtained by full-potential linearized
augmented plane wave (FLAPW) energy band calculations.
The splitting of energy bands due to spin-orbit coupling is
clearly seen in both figures.
These figures exhibit some characteristic features of the band structure
discussed above:
the Weyl points on the $\Gamma$-X line, the two- and four-fold
degeneracies at the $\Gamma$ point, the two- and six-fold degeneracies
at the R point, and the Weyl points on the $\Gamma$-R line.


\section{Space groups No.~61 and No.~205}
\label{sec: 205}

The SG61 ($Pbca$) and SG205 ($P$a-3) are obtained from
the SG19 and SG198, respectively, by including
inversion symmetry denoted by $P$.
In the presence of both inversion and time-reversal symmetries,
Bloch states form Kramers doublets at every $\bm{k}$ point
in the Brillouin zone, because
$P\Theta$ preserves $\bm{k}$ and satisfies
$(P\Theta)^2=-1$.

\subsection{Space group No.~61}

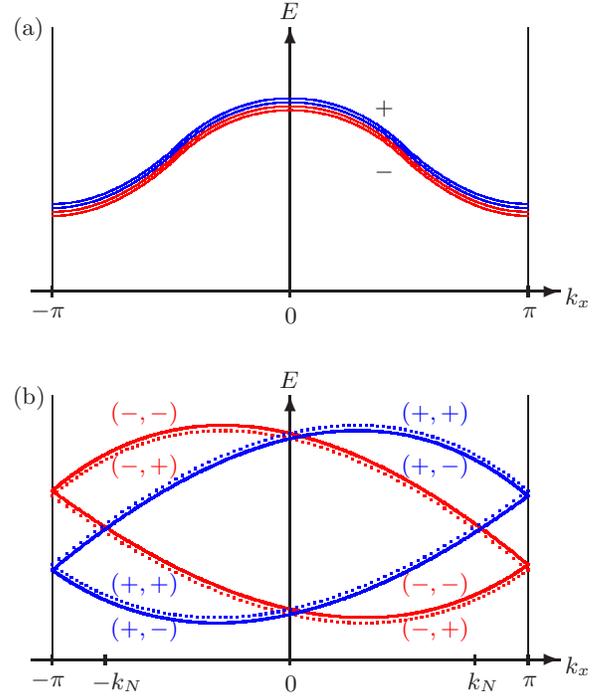
\begin{figure}
\begin{center}
%
%
\begin{picture}(250,130)
\put(13,117){(a)}
\put(28,20){\line(0,1){100}}
\put(208,20){\line(0,1){100}}
\thicklines
\put(118,20){\vector(0,1){100}}
\put(114,123){$E$}
\put(20,20){\vector(1,0){200}}
\put(222,15){$k_x$}
\put(28,18){\line(0,1){4}}
\put(20,10){$-\pi$}
\put(118,18){\line(0,1){4}}
\put(116,8){$0$}
\put(208,18){\line(0,1){4}}
\put(206,10){$\pi$}
\put(150,87){$+$}
\put(150,63){$-$}
\thinlines
\color{red}{
\qbezier(28,50)(50,50)(73,70)}
\qbezier(73,70)(93,90)(118,90)
\qbezier(28,48.5)(50,48.5)(73,68.5)
\qbezier(73,68.5)(93,88.5)(118,88.5)
\qbezier(118,90)(143,90)(163,70)
\qbezier(163,70)(186,50)(208,50)
\qbezier(118,88.5)(143,88.5)(163,68.5)
\qbezier(163,68.5)(186,48.5)(208,48.5)
\color{blue}{
\qbezier(28,51.5)(50,51.5)(73,71.5)}
\qbezier(73,71.5)(93,91.5)(118,91.5)
\qbezier(28,53)(50,53)(73,73)
\qbezier(73,73)(93,93)(118,93)
\qbezier(118,91.5)(143,91.5)(163,71.5)
\qbezier(163,71.5)(186,51.5)(208,51.5)
\qbezier(118,93)(143,93)(163,73)
\qbezier(163,73)(186,53)(208,53)
\end{picture}
\vskip 3mm
%
\begin{picture}(250,125)(0,5)
\put(13,117){(b)}
\put(28,20){\line(0,1){100}}
\put(208,20){\line(0,1){100}}
\thicklines
\put(118,20){\vector(0,1){100}}
\put(114,123){$E$}
\put(20,20){\vector(1,0){200}}
\put(222,15){$k_x$}
\put(28,18){\line(0,1){4}}
\put(20,10){$-\pi$}
\put(118,18){\line(0,1){4}}
\put(116,8){$0$}
\put(208,18){\line(0,1){4}}
\put(206,10){$\pi$}
\put(188,18){\line(0,1){4}}
\put(185,9){$k_N$}
\put(48,18){\line(0,1){4}}
\put(43,9){$-k_N$}
\color{red}
\qbezier(28,84)(100,145)(208,56)
\qbezier(28,84)(136,5)(208,56)
\qbezier[80](28,82)(100,143)(208,54)
\qbezier[80](28,82)(136,3)(208,54)
\put(50,111){$(-,-)$}
\put(50,91){$(-,+)$}
\put(160,46){$(-,-)$}
\put(160,29){$(-,+)$}
\color{blue}
\qbezier[80](28,56)(100,5)(208,84)
\qbezier[80](28,56)(136,145)(208,84)
\qbezier(28,54)(100,3)(208,82)
\qbezier(28,54)(136,143)(208,82)
\put(160,111){$(+,+)$}
\put(160,91){$(+,-)$}
\put(50,46){$(+,+)$}
\put(50,29){$(+,-)$}
\end{picture}
\end{center}
\vskip -3mm
\caption{
Schematic band structures for SG61 and SG205
(a) along the invariant line $(k_x,\pi,0)$
and (b) along the invariant line $(k_x,0,\pi)$.
In (a) the energy bands are four-fold degenerate:
two bands with the $\widetilde{C}_{2x}$-eigenvalue $+ie^{-ik_x/2}$
and the other two bands with
$-ie^{-ik_x/2}$.
In (b)
the energy bands labeled by $(s_1,s_2)$ have the eigenvalues
$(G_z,\widetilde{C}_{2x})=(s_1,s_2)ie^{ik_x/2}$, where $s_{1,2}=\pm$.
The red (blue) curves represent bands with $s_1=+$ ($-$), while
the solid (dashed) curves represent bands with $s_2=-$ ($+$),
respectively.
}
\label{fig: bands with inversion}
\end{figure}

\begin{figure}
\begin{center}
\begin{picture}(200,175)(0,5)
\put(78,39){0}
\put(80,50){\vector(-2,-1){78}}
\put(-8,8){$k_x$}
\put(18,11){$\pi$}
\put(80,50){\vector(1,0){122}}
\put(77,175){$k_z$}
\put(70,150){$\pi$}
\put(80,50){\vector(0,1){120}}
\put(200,40){$k_y$}
\put(177,41){$\pi$}
\put(70,50){$\Gamma$}
\put(10,21){X}
\put(0,123){U$\,$(M)}
\put(83,154){Z$\,$(X)}
\put(119,10){S$\,$(M)}
\put(182.5,53){Y$\,$(X)}
\put(181,153){T$\,$(M)}
\put(117,125){R}
\thicklines
\put(20,20){\line(1,0){100}}
\put(120,20){\line(0,1){100}}
\put(20,120){\line(1,0){100}}
\put(20,120){\line(2,1){60}}
\put(120,120){\line(2,1){60}}
\put(180,50){\line(0,1){100}}
%
\color{red}
\put(20,20){\line(0,1){100}}
\put(80,150){\line(1,0){100}}
\put(120,20){\line(2,1){60}}
\color{red}
\qbezier(40,130)(80,130)(120,120)
\qbezier(86,20)(86,60)(120,120)
\qbezier(180,116)(155,100)(120,120)
\end{picture}
\end{center}
\caption{
Four-fold degenerate line nodes on the boundaries of
the Brillouin zone for SG61 (SG205).
Here only one eighth of the Brillouin zone ($0\le k_\alpha\le\pi$) is shown.
}
\label{fig: BZ/8 61}
\end{figure}
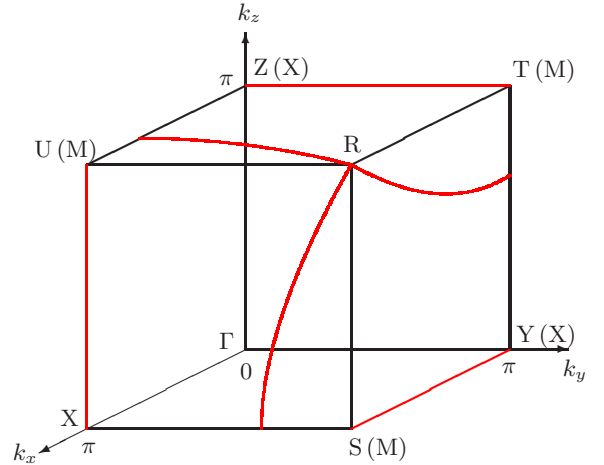

The inversion operator $P$ and the screw rotations $\widetilde{C}_{2\alpha}$
($\alpha=x,y,z$) obey the commutation relations
\begin{equation}
\widetilde{C}_{2x}P=T_{(1,1,0)}P\widetilde{C}_{2x}
=e^{-i(k_x-k_y)}P\widetilde{C}_{2x},
\label{C_2x vs P}
\end{equation}
and its cyclic permutations about $(x,y,z)$.
It follows from Eq.\ (\ref{C_2x vs P}) that
\begin{eqnarray}
\widetilde{C}_{2x}P|\pm,n,\bm{k}\rangle
&=&e^{-i(k_x-k_y)}P\widetilde{C}_{2x}|\pm,n,\bm{k}\rangle
\nonumber\\
&=&
\pm ie^{-\frac{i}{2}k_x+ik_y}P|\pm,n,\bm{k}\rangle,
\end{eqnarray}
where $|\pm,n,\bm{k}\rangle$ are the Bloch states
satisfying Eq.\ (\ref{C_2x | >})
and $\bm{k}$ is on an invariant line of $\widetilde{C}_{2x}$.
Applying $\Theta$ yields
\begin{equation}
\widetilde{C}_{2x}P\Theta|\pm,n,\bm{k}\rangle
=\mp ie^{\frac{i}{2}k_x-ik_y}P\Theta|\pm,n,\bm{k}\rangle,
\label{C_2x and PTheta}
\end{equation}
from which we deduce that the band structure of the SG61
is changed from that of the SG19 as follows.
The band structure along the $\Gamma$-X line $(k_x,0,0)$
is changed from the one shown in Fig.~\ref{fig: basic bands}(a)
into Fig.~\ref{fig: basic bands}(c).
On the Y-S line $(k_x,\pi,0)$,
a Kramers-degenerate pair 
$|s,n,\bm{k}\rangle$ and $P\Theta|s,n,\bm{k}\rangle$
have the same $\widetilde{C}_{2x}$-eigenvalue
$ise^{ik_x/2}$ ($s=\pm$), whereas $|s,n,\bm{k}\rangle$ and
$\Theta\widetilde{C}_{2y}|s,n,\bm{k}\rangle$ have different
$\widetilde{C}_{2x}$-eigenvalues.
Thus, the band structure is changed from Fig.~\ref{fig: basic bands}(c)
to Fig.~\ref{fig: bands with inversion}(a), and the energy bands
form a four-fold degenerate Dirac line node.
Repeating the same consideration for invariant lines of
$\widetilde{C}_{2y}$ and $\widetilde{C}_{2z}$, we conclude that
all the energy bands along the X-U, Y-S, and Z-T lines are
four-fold degenerate nodal lines;
see Fig.~\ref{fig: BZ/8 61}.
While the band structure on the T-R line $(k_x,\pi,\pi)$
remains qualitatively the same as in Fig.~\ref{fig: basic bands}(b),
the connection of bands at the R point is modified
from Fig.~\ref{fig: connection}(b),
as we discuss below.

Let us consider glide mirror transformation defined by
$G_z=\widetilde{C}_{2z}P$, which acts as
\begin{equation}
G_z :
(x,y,z)\rightarrow\textstyle{(x+\frac12,y,-z+\frac12)}
\label{G_z}
\end{equation}
on the real-space coordinates and as $i\sigma_z$ in the spin space.
On the $k_z=\pi$ plane, which is invariant under $G_z$,
Bloch states are chosen as eigenstates of $G_z$ such that
\begin{equation}
G_z|\pm,n,\bm{k}\rangle_g=\pm ie^{ik_x/2}|\pm,n,\bm{k}\rangle_g,
\label{G_z Bloch}
\end{equation}
where $n$ is a band index and $\bm{k}=(k_x,k_y,\pi)$.
The inversion $P$ and the glide mirror $G_z$ satisfy the algebra
\begin{equation}
G_zP=T_{(1,0,1)}PG_z=e^{-i(k_x-k_z)}PG_z
\label{G_z vs P}
\end{equation}
when acting on Bloch states.
It follows that
\begin{equation}
G_zP\Theta|\pm,n,\bm{k}\rangle_g
=\pm ie^{ik_x/2}P\Theta|\pm,n,\bm{k}\rangle_g
\label{G_z and PTheta}
\end{equation}
at $k_z=\pi$.
Thus, a pair of Kramers-degenerate states, $|s,n,\bm{k}\rangle_g$ and
$P\Theta|s,n,\bm{k}\rangle_g$, have the same $G_z$-eigenvalues
$ise^{ik_x/2}$ ($s=+$ or $-$) on the $k_z=\pi$ plane.

The screw rotation $\widetilde{C}_{2y}$ and the glide mirror $G_z$
satisfy the algebra
\begin{equation}
G_z\widetilde{C}_{2y}=-T_{(1,0,0)}\widetilde{C}_{2y}G_z
=-e^{-ik_x}\widetilde{C}_{2y}G_z,
\label{G_z vs C_2y}
\end{equation}
where the minus signs are due to the anticommutation
relation $\{\sigma_z,\sigma_y\}=0$.
It then follows that
\begin{equation}
G_z\widetilde{C}_{2y}|\pm,n,\bm{k}\rangle_g
=\mp ie^{-ik_x/2}\widetilde{C}_{2y}|\pm,n,\bm{k}\rangle_g.
\end{equation}
The $G_z$-eigenvalue of the Bloch state
$\widetilde{C}_{2y}|s,n,\bm{k}\rangle_g$
is thus different from that of $|s,n,\bm{k}\rangle_g$.
This implies that two-fold degenerate bands with $G_z=+ie^{ik_x/2}$
and two-fold degenerate bands with $G_z=-ie^{ik_x/2}$ must cross
along the Z-T line $(0,k_y,\pi)$.
Hence the energy bands are four-fold degenerate along the Z-T line
in agreement with the discussion above.

Similarly to Eq.\ (\ref{G_z vs C_2y}), $G_z$ and $\widetilde{C}_{2x}$
satisfy the algebra
\begin{equation}
G_z\widetilde{C}_{2x}=-T_{(0,0,1)}\widetilde{C}_{2x}G_z
=-e^{ik_z}\widetilde{C}_{2x}G_z,
\label{G_z vs C_2x}
\end{equation}
and, in particular, they commute at $k_z=\pi$.
This means that Bloch states can be chosen to be eigenstates of
both $G_z$ and $\widetilde{C}_{2x}$ on the Z-U line;
\begin{equation}
(G_z,\widetilde{C}_{2x})|s_1,s_2,n,\bm{k}\rangle
=(s_1,s_2)ie^{ik_x/2}|s_2,s_2,n,\bm{k}\rangle,
\end{equation}
where $s_{1,2}=\pm$, $n$ is a band index, and
$\bm{k}=(k_x,0,\pi)$.
We find from Eqs.\ (\ref{C_2x and PTheta}) and (\ref{G_z and PTheta})
that the $P\Theta$ transformation changes the eigenvalues,
\begin{equation}
(G_z,\widetilde{C}_{2x})\stackrel{P\Theta}{\longrightarrow}
(G_z,-\widetilde{C}_{2x}).
\end{equation}
We also notice that
\begin{equation}
(G_z,\widetilde{C}_{2x})\stackrel{\Theta}{\longrightarrow}
(-G_z,-\widetilde{C}_{2x})
\end{equation}
at $\bm{k}=(0,0,\pi)$, and
\begin{equation}
(G_z,\widetilde{C}_{2x})\stackrel{\Theta}{\longrightarrow}
(G_z,\widetilde{C}_{2x})
\end{equation}
at $\bm{k}=(\pi,0,\pi)$.
These three relations lead us to conclude that
the energy levels are four-fold degenerate
at $\bm{k}=(0,0,\pi)$ and $(\pi,0,\pi)$,
but degenerate states exchange their partners between these $\bm{k}$ points,
as shown schematically
in Fig.~\ref{fig: bands with inversion}(b).
It is important to note that there are four-fold degenerate band crossing
points at $k_x=\pm k_N$ besides those at $k_x=0,\pm\pi$.

\begin{figure}
\begin{center}
\begin{picture}(250,195)(-10,-33)
\thicklines
\put(28,20){\line(0,1){90}}
\put(13,8){U$(\pi,0,\pi)$}
\put(178,80){\line(0,1){80}}
\put(176,71){T$(0,\pi,\pi)$}
\put(180,-22){\line(0,1){85}}
\put(146.5,-28){S$(\pi,\pi,0)$}
\put(118,20){\vector(0,1){100}}
\put(114,123){$E$}
\put(92,7.5){R$(\pi,\pi,\pi)$}
\put(118,20){\vector(-1,0){100}}
\put(1,18){$-k_y$}
\put(118,20){\vector(1,1){74}}
\put(194,96){$-k_x$}
\put(118,20){\vector(3,-2){70}}
\put(192,-30){$-k_z$}
%
%
%
%
\color{blue}
\qbezier(28,82)(80,102)(118,102)
\qbezier(28,80)(80,100)(118,100)
\qbezier(28,50)(80,67)(118,67)
\qbezier(28,48)(80,65)(118,65)
\color{red}
\qbezier(28,82)(80,67)(118,67)
\qbezier(28,80)(80,65)(118,65)
\qbezier(28,50)(80,35)(118,35)
\qbezier(28,48)(80,33)(118,33)
%
\color{red}
\qbezier[30](118,67)(140,78)(178,130)
\qbezier[30](118,65)(140,76)(178,128)
\qbezier[40](118,35)(138,47)(178,100)
\qbezier[40](118,33)(140,45)(178,98)
\color{blue}
\qbezier[30](118,102)(140,120)(178,130)
\qbezier[30](118,100)(140,118)(178,128)
\qbezier[35](118,67)(138,80)(178,100)
\qbezier[35](118,65)(138,78)(178,98)
%
\color{red}
\qbezier(118,102)(142,96)(180,40)
\qbezier(118,100)(141,94)(180,38)
\qbezier(118,35)(140,12)(180,10)
\qbezier(118,33)(140,10)(180,8)
\color{blue}
\qbezier(118,67)(140,54)(180,40)
\qbezier(118,65)(140,52)(180,38)
\qbezier(118,67)(140,50)(180,10)
\qbezier(118,65)(139,48)(180,8)
\end{picture}
\end{center}
\caption{
Band dispersion along the R-S, R-T, and R-U lines
for SG61.
The energy bands with blue (red) color have the $\widetilde{C}_{2\alpha}$
eigenvalue $+ie^{ik_\alpha/2}$ ($-ie^{ik_\alpha/2}$),
where $\alpha=x$, $y$, or $z$.
}
\label{fig: connection at R 61}
\end{figure}
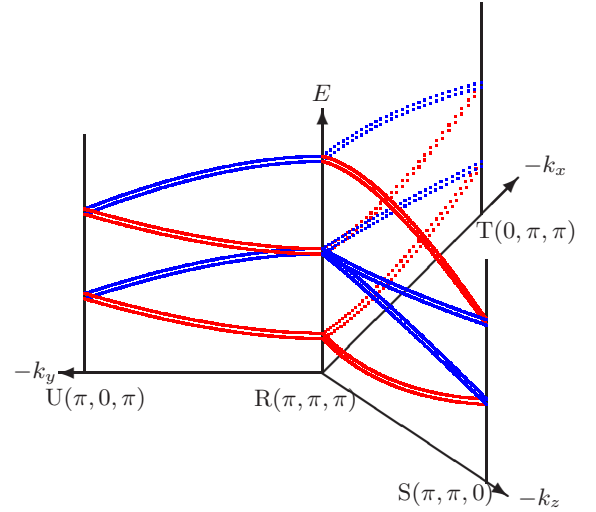

As we can see from Fig.~\ref{fig: bands with inversion}(b),
the four-fold degenerate energy levels at the U point $\bm{k}=(\pi,0,\pi)$
have the same $G_z$ eigenvalue ($+1$ or $-1$).
On the other hand, the four-fold degenerate energy levels
on the Z-T line $(0,k_y,\pi)$ consist of two states with
$G_z=+i$ and two states with $G_z=-i$.
Since the $k_x$-dependent eigenvalues $G_z=\pm ie^{ik_x/2}$ are
well defined on the whole $k_z=\pi$ plane,
any curve connecting a point on the Z-T line and
the U point must have a four-fold degenerate crossing point on the curve,
like the band crossing at $k_x=k_N$ on the Z-U line
in Fig.~\ref{fig: bands with inversion}(b).
Such crossing points form a four-fold degenerate nodal line
on the $k_z=\pi$ plane.
(A similar mechanism of forming line nodes was previously discussed
in Refs.~\onlinecite{Chen2016,Bzdusek2016}.)
The same argument can be applied to the $k_x=\pi$ plane and the $k_y=\pi$ plane
by considering $G_x$ and $G_z$ eigenvalues, respectively.
These nodal lines should be connected at the common edges of the three planes,
i.e., at the R point $(\pi,\pi,\pi)$; see Fig.~\ref{fig: BZ/8 61}.
The energy band structure along the R-S, R-T, and R-U lines are
schematically shown in Fig.~\ref{fig: connection at R 61}.
Among the eight bands drawn in Fig.~\ref{fig: connection at R 61},
the middle four bands are connected at the R point,
while the higher/lower four bands are connected at the S, T, and
U points.
We notice that the inversion symmetry forces a four-fold degenerate
point to be fixed at the R point, while it can be on the R-S,
R-T, or R-U line for the SG19.

\subsection{Space group No.~205}

The SG205 has the additional $C_3$ symmetry compared with 
the SG61.

\begin{figure}
\begin{center}
\begin{picture}(200,250)(0,0)
\thicklines
\put(10,10){\line(1,0){180}}
\put(8,0){$\Gamma$}
\put(187,0){R}
\put(10,10){\vector(0,1){235}}
\put(0,242){$E$}
\put(190,10){\line(0,1){235}}
\put(2,26){4}
\put(2,33){2}
\put(2,49){2}
\put(2,72){4}
\put(2,111){2}
\put(2,123){4}
\put(2,137){4}
\put(2,197){2}
\put(193,33){2}
\put(193,50){6}
\put(193,102){2}
\put(193,109){6}
\put(193,188){6}
\put(193,198){2}
\color{red}
\qbezier(10,30)(30,30)(55,22)
\qbezier(100,14)(80,14)(55,22)
\qbezier(100,14)(120,14)(145,25)
\qbezier(190,36)(170,36)(145,25)
\qbezier(10,36)(40,36)(60,30)
\qbezier(110,24)(80,24)(60,30)
\qbezier(110,24)(125,24)(150,38)
\qbezier(190,51.5)(175,51.5)(150,38)
\qbezier(10,52)(20,52)(40,60)
\qbezier(40,60)(60,66)(70,66)
\qbezier(70,66)(100,66)(130,60)
\qbezier(190,52.5)(160,52.5)(130,60)
\qbezier(10,74.5)(20,75)(30,72)
\qbezier(30,72)(40,69)(50,69)
\qbezier(50,69)(70,68)(120,87)
\qbezier(190,105)(170,105)(120,87)
\qbezier(10,114)(15,114)(20,112)
\qbezier(20,112)(25,110)(30,110)
\qbezier(30,110)(35,110)(40,115)
\qbezier(40,115)(45,120)(50,120)
\qbezier(50,120)(60,120)(70,110)
\qbezier(70,110)(80,100)(90,100)
\qbezier(90,100)(105,100)(140,106)
\qbezier(190,110.5)(160,111)(140,106)
\qbezier(10,127)(17,127)(24,132)
\qbezier(24,132)(30,137)(35,137)
\qbezier(35,137)(38,137)(43,133)
\qbezier(43,133)(48,130)(51,130)
\qbezier(51,130)(75,130)(110,155)
\qbezier(110,155)(135,180)(150,180)
\qbezier(150,180)(160,180)(170,145)
\qbezier(170,145)(180,111)(190,111)
\qbezier(10,140)(20,140)(38.5,163)
\qbezier(38.5,163)(55,187)(65,187)
\qbezier(65,187)(80,187)(100,195)
\qbezier(100,195)(120,203)(135,203)
\qbezier(135,203)(140,203)(150,195)
\qbezier(150,195)(157,185)(164,184)
\qbezier(177,187)(171,184)(164,184)
\qbezier(190,189.5)(183,189.5)(177,187)
\qbezier(10,200)(14,200)(18,184)
\qbezier(18,184)(22,168)(26,168)
\qbezier(26,168)(30,168)(44,187)
\qbezier(44,187)(55,200)(62,205)
\qbezier(62,205)(70,210)(85,225)
\qbezier(85,225)(100,240)(115,236)
\qbezier(115,236)(120,233)(132,219)
\qbezier(132,219)(145,205)(154,206)
\qbezier(154,206)(163,205)(172,198)
\qbezier(172,198)(183,191)(190,190.5)
\color{blue}
\qbezier(10,30.5)(30,31)(55,27)
\qbezier(100,23)(80,23)(55,27)
\qbezier(100,23)(120,23)(145,38)
\qbezier(190,52)(170,52)(145,38)
\qbezier(10,75)(50,75)(100,88)
\qbezier(100,88)(160,109)(190,110)
\qbezier(10,126.5)(15,127)(20,124)
\qbezier(20,124)(25,121)(30,121)
\qbezier(30,121)(45,121)(80,155)
\qbezier(80,155)(100,175)(120,180)
\qbezier(120,180)(160,190)(190,190)
\qbezier(10,140.5)(20,141)(37,165)
\qbezier(37,165)(52,190)(62,192)
\qbezier(62,192)(75,193)(97,200)
\qbezier(97,200)(115,207)(125,210)
\qbezier(125,210)(133,214)(140,215)
\qbezier(140,215)(147,220)(165,208)
\qbezier(165,208)(182,199)(190,200)
\end{picture}
\end{center}
\caption{
Schematic band structure along the $C_3$-invariant $\Gamma$-R line
for SG205.
In this figure each band is two-fold degenerate because of the
$PT$ symmetry.
The red curves represent two bands with $C_3$ eigenvalues equal to
$\exp(\pm i\pi/3)$,
and the blue curves represent two bands with $C_3=-1$.
The integer numbers next to the vertical axes show the degeneracy of energy
levels at the $\Gamma$ and R points.
}
\label{fig: Gamma-R}
\end{figure}
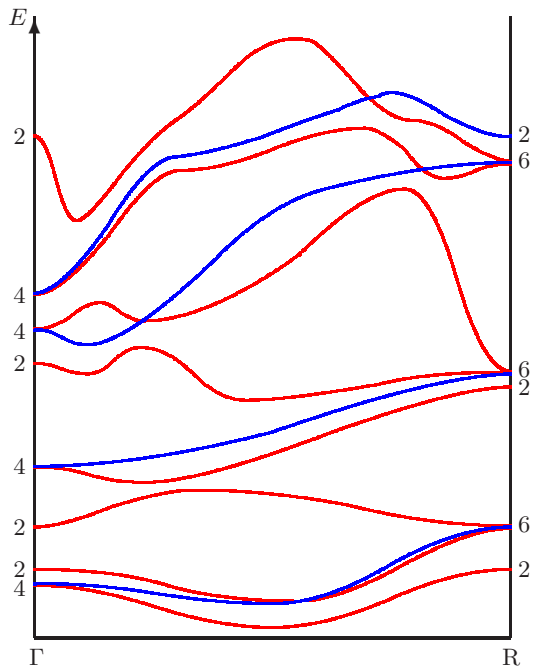

Along the $C_3$-invariant $\Gamma$-R line,
the energy bands are classified in terms of the $C_3$ eigenvalues.
The Kramers theorem from the $TP$ symmetry implies that doubly
degenerate bands along the $\Gamma$-R line have a pair of
$C_3$-eigenvalues $(-1,-1)$ or $(e^{i\pi/3},e^{-i\pi/3})$.
Energy bands with different pairs of $C_3$-eigenvalues
can cross to form four-fold degenerate Dirac points \cite{WengReview,BJYang}.
As in the SG 198,
the energy levels at the $\Gamma$ point are either doublets with
the $C_3$-eigenvalues $(e^{i\pi/3},e^{-i\pi/3})$ or
quartets with $(-1,e^{i\pi/3},e^{-i\pi/3},-1)$,
while those at the R point are either doublets
with $C_3=(-1,-1),(e^{i\pi/3},e^{-i\pi/3})$ or sextuplets
with $C_3=(-1,-1,e^{i\pi/3},e^{i\pi/3},e^{-i\pi/3},e^{-i\pi/3})$.
Figure~\ref{fig: Gamma-R} shows an example of the energy band structure
along the $\Gamma$-R line, which can be compared with
the energy bands of CoSe$_2$ \cite{Teruya2016}.
In Fig.~\ref{fig: Gamma-R} curves represent doubly degenerate energy bands
with a pair of $C_3$ eigenvalues $(e^{i\pi/3},e^{-i\pi/3})$ or $(-1,-1)$,
and each crossing of two curves with different $C_3$ eigenvalues
realizes a Dirac point. 
The lowest eight bands and the higher sixteen bands
form two independent groups of connected energy bands
in Fig.~\ref{fig: Gamma-R}.

The band structure along the R-M line in the SG205 is qualitatively
the same as Fig.~\ref{fig: connection at R with C_3}.
Comparing it with Fig.~\ref{fig: connection at R 61} for the SG61,
we see that the $C_3$ symmetry promotes the four-fold degeneracy
to the six-fold degeneracy at the R point.

Our argument for the four-fold degenerate line nodes in the SG61
can be applied to the SG205 as well.
Therefore, the energy bands should be four-fold degenerate
along the X-M line (corresponding to the S-Y, T-Z, and U-X lines
of the SG61),
and additional four-fold degenerate line nodes
connecting at the R point should appear as shown in
Fig.~\ref{fig: BZ/8 61}.

CoSe$_2$ has crystal structure of the SG205, and its energy band
structure computed with the FLAPW method
is shown in Fig.~7 of Ref.~\onlinecite{Teruya2016}.
The computed energy band structure has the characteristic features
we discussed above.
The energy bands are four-fold degenerate along one of the X-M line
corresponding to the S-Y, T-Z, U-X lines
(note that there are two inequivalent X-M lines in the chiral crystal
structure of the SG205).
On the other X-M line, a four-fold degenerate band crossing point
is present near the M point, which should be a part of the four-fold
degenerate Dirac line node connecting the R point.
The four-fold degenerate bands are dispersive and cross the Fermi energy.
Unfortunately, the full shape of the line node connecting the R point
is not elucidated in Ref.~\onlinecite{Teruya2016}
because the bands are computed only along high-symmetry lines.
Furthermore, quartets at the $\Gamma$ point and sextuplets at the R points
are clearly observed, and there are several Dirac points on the
$C_3$-symmetric $\Gamma$-R line.

\section{Summary}

In this paper we have discussed various types of band degeneracies
caused by nonsymmorphic crystal symmetries
such as screw rotation and glide mirror.

It was shown that a screw rotation symmetry necessarily leads to
the formation of Weyl points (Fig.~\ref{fig: basic bands}(a))
and the presence of multiple screw rotation symmetries with orthogonal
rotation axes generates double Weyl points on the Brillouin zone boundary.
These Weyl points produce multiple surface Fermi arcs, which can
in principle be observed with angle-resolved photoemission spectroscopy.
The Weyl points also affect transport properties and give rise
to such anomalous effects as negative magnetoresistance
due to chiral anomaly \cite{Son}.

An additional inversion symmetry promotes (double) Weyl
points to four-fold degenerate line nodes,
which will give rise to ``drumhead'' surface states \cite{Chan2016}.
Furthermore, dispersive line nodes can yield
Fermi surfaces touching at points on the line nodes.


\acknowledgments
This work was in part supported by JSPS Kakenhi (No.~15K05141)
from Japan Society for the Promotion of Science.
The author is grateful to T. Morimoto, K. Shiozaki, H. Watanabe,
and B.-J. Yang for helpful discussions.

\bibliography{semimetals}   

\begin{thebibliography}{30}
\expandafter\ifx\csname natexlab\endcsname\relax\def\natexlab#1{#1}\fi
\expandafter\ifx\csname bibnamefont\endcsname\relax
  \def\bibnamefont#1{#1}\fi
\expandafter\ifx\csname bibfnamefont\endcsname\relax
  \def\bibfnamefont#1{#1}\fi
\expandafter\ifx\csname citenamefont\endcsname\relax
  \def\citenamefont#1{#1}\fi
\expandafter\ifx\csname url\endcsname\relax
  \def\url#1{\texttt{#1}}\fi
\expandafter\ifx\csname urlprefix\endcsname\relax\def\urlprefix{URL }\fi
\providecommand{\bibinfo}[2]{#2}
\providecommand{\eprint}[2][]{\url{#2}}

\bibitem[{\citenamefont{Hasan and Kane}(2010)}]{HasanKane}
\bibinfo{author}{\bibfnamefont{M.~Z.} \bibnamefont{Hasan}} \bibnamefont{and}
  \bibinfo{author}{\bibfnamefont{C.~L.} \bibnamefont{Kane}},
  \bibinfo{journal}{Rev. Mod. Phys.} \textbf{\bibinfo{volume}{82}},
  \bibinfo{pages}{3045} (\bibinfo{year}{2010}).

\bibitem[{\citenamefont{Qi and Zhang}(2011)}]{QiZhang}
\bibinfo{author}{\bibfnamefont{X.-L.} \bibnamefont{Qi}} \bibnamefont{and}
  \bibinfo{author}{\bibfnamefont{S.-C.} \bibnamefont{Zhang}},
  \bibinfo{journal}{Rev. Mod. Phys.} \textbf{\bibinfo{volume}{83}},
  \bibinfo{pages}{1057} (\bibinfo{year}{2011}).

\bibitem[{\citenamefont{Murakami}(2007)}]{Murakami}
\bibinfo{author}{\bibfnamefont{S.}~\bibnamefont{Murakami}},
  \bibinfo{journal}{New Journal of Physics} \textbf{\bibinfo{volume}{9}},
  \bibinfo{pages}{356} (\bibinfo{year}{2007}).

\bibitem[{\citenamefont{Wan et~al.}(2011)\citenamefont{Wan, Turner, Vishwanath,
  and Savrasov}}]{WanTurner}
\bibinfo{author}{\bibfnamefont{X.}~\bibnamefont{Wan}},
  \bibinfo{author}{\bibfnamefont{A.~M.} \bibnamefont{Turner}},
  \bibinfo{author}{\bibfnamefont{A.}~\bibnamefont{Vishwanath}},
  \bibnamefont{and} \bibinfo{author}{\bibfnamefont{S.~Y.}
  \bibnamefont{Savrasov}}, \bibinfo{journal}{Phys. Rev. B}
  \textbf{\bibinfo{volume}{83}}, \bibinfo{pages}{205101}
  (\bibinfo{year}{2011}).

\bibitem[{\citenamefont{Young et~al.}(2012)\citenamefont{Young, Zaheer, Teo,
  Kane, Mele, and Rappe}}]{Young2012}
\bibinfo{author}{\bibfnamefont{S.~M.} \bibnamefont{Young}},
  \bibinfo{author}{\bibfnamefont{S.}~\bibnamefont{Zaheer}},
  \bibinfo{author}{\bibfnamefont{J.~C.~Y.} \bibnamefont{Teo}},
  \bibinfo{author}{\bibfnamefont{C.~L.} \bibnamefont{Kane}},
  \bibinfo{author}{\bibfnamefont{E.~J.} \bibnamefont{Mele}}, \bibnamefont{and}
  \bibinfo{author}{\bibfnamefont{A.~M.} \bibnamefont{Rappe}},
  \bibinfo{journal}{Phys. Rev. Lett.} \textbf{\bibinfo{volume}{108}},
  \bibinfo{pages}{140405} (\bibinfo{year}{2012}).

\bibitem[{\citenamefont{Weng et~al.}(2016)\citenamefont{Weng, Dai, and
  Fang}}]{WengReview}
\bibinfo{author}{\bibfnamefont{H.}~\bibnamefont{Weng}},
  \bibinfo{author}{\bibfnamefont{X.}~\bibnamefont{Dai}}, \bibnamefont{and}
  \bibinfo{author}{\bibfnamefont{Z.}~\bibnamefont{Fang}},
  \bibinfo{journal}{Journal of Physics: Condensed Matter}
  \textbf{\bibinfo{volume}{28}}, \bibinfo{pages}{303001}
  (\bibinfo{year}{2016}).

\bibitem[{\citenamefont{Wang et~al.}(2012)\citenamefont{Wang, Sun, Chen,
  Franchini, Xu, Weng, Dai, and Fang}}]{Wang2012}
\bibinfo{author}{\bibfnamefont{Z.}~\bibnamefont{Wang}},
  \bibinfo{author}{\bibfnamefont{Y.}~\bibnamefont{Sun}},
  \bibinfo{author}{\bibfnamefont{X.-Q.} \bibnamefont{Chen}},
  \bibinfo{author}{\bibfnamefont{C.}~\bibnamefont{Franchini}},
  \bibinfo{author}{\bibfnamefont{G.}~\bibnamefont{Xu}},
  \bibinfo{author}{\bibfnamefont{H.}~\bibnamefont{Weng}},
  \bibinfo{author}{\bibfnamefont{X.}~\bibnamefont{Dai}}, \bibnamefont{and}
  \bibinfo{author}{\bibfnamefont{Z.}~\bibnamefont{Fang}},
  \bibinfo{journal}{Phys. Rev. B} \textbf{\bibinfo{volume}{85}},
  \bibinfo{pages}{195320} (\bibinfo{year}{2012}).

\bibitem[{\citenamefont{Wang et~al.}(2013)\citenamefont{Wang, Weng, Wu, Dai,
  and Fang}}]{Wang2013}
\bibinfo{author}{\bibfnamefont{Z.}~\bibnamefont{Wang}},
  \bibinfo{author}{\bibfnamefont{H.}~\bibnamefont{Weng}},
  \bibinfo{author}{\bibfnamefont{Q.}~\bibnamefont{Wu}},
  \bibinfo{author}{\bibfnamefont{X.}~\bibnamefont{Dai}}, \bibnamefont{and}
  \bibinfo{author}{\bibfnamefont{Z.}~\bibnamefont{Fang}},
  \bibinfo{journal}{Phys. Rev. B} \textbf{\bibinfo{volume}{88}},
  \bibinfo{pages}{125427} (\bibinfo{year}{2013}).

\bibitem[{\citenamefont{Lv et~al.}(2015)\citenamefont{Lv, Weng, Fu, Wang, Miao,
  Ma, Richard, Huang, Zhao, Chen et~al.}}]{Lv2015}
\bibinfo{author}{\bibfnamefont{B.~Q.} \bibnamefont{Lv}},
  \bibinfo{author}{\bibfnamefont{H.~M.} \bibnamefont{Weng}},
  \bibinfo{author}{\bibfnamefont{B.~B.} \bibnamefont{Fu}},
  \bibinfo{author}{\bibfnamefont{X.~P.} \bibnamefont{Wang}},
  \bibinfo{author}{\bibfnamefont{H.}~\bibnamefont{Miao}},
  \bibinfo{author}{\bibfnamefont{J.}~\bibnamefont{Ma}},
  \bibinfo{author}{\bibfnamefont{P.}~\bibnamefont{Richard}},
  \bibinfo{author}{\bibfnamefont{X.~C.} \bibnamefont{Huang}},
  \bibinfo{author}{\bibfnamefont{L.~X.} \bibnamefont{Zhao}},
  \bibinfo{author}{\bibfnamefont{G.~F.} \bibnamefont{Chen}},
  \bibnamefont{et~al.}, \bibinfo{journal}{Phys. Rev. X}
  \textbf{\bibinfo{volume}{5}}, \bibinfo{pages}{031013} (\bibinfo{year}{2015}).

\bibitem[{\citenamefont{Xu et~al.}(2015)\citenamefont{Xu, Belopolski, Alidoust,
  Neupane, Bian, Zhang, Sankar, Chang, Yuan, Lee et~al.}}]{Xu2015}
\bibinfo{author}{\bibfnamefont{S.-Y.} \bibnamefont{Xu}},
  \bibinfo{author}{\bibfnamefont{I.}~\bibnamefont{Belopolski}},
  \bibinfo{author}{\bibfnamefont{N.}~\bibnamefont{Alidoust}},
  \bibinfo{author}{\bibfnamefont{M.}~\bibnamefont{Neupane}},
  \bibinfo{author}{\bibfnamefont{G.}~\bibnamefont{Bian}},
  \bibinfo{author}{\bibfnamefont{C.}~\bibnamefont{Zhang}},
  \bibinfo{author}{\bibfnamefont{R.}~\bibnamefont{Sankar}},
  \bibinfo{author}{\bibfnamefont{G.}~\bibnamefont{Chang}},
  \bibinfo{author}{\bibfnamefont{Z.}~\bibnamefont{Yuan}},
  \bibinfo{author}{\bibfnamefont{C.-C.} \bibnamefont{Lee}},
  \bibnamefont{et~al.}, \bibinfo{journal}{Science}
  \textbf{\bibinfo{volume}{349}}, \bibinfo{pages}{613} (\bibinfo{year}{2015}).

\bibitem[{\citenamefont{Parameswaran et~al.}(2013)\citenamefont{Parameswaran,
  Turner, Arovas, and Vishwanath}}]{Sid}
\bibinfo{author}{\bibfnamefont{S.~A.} \bibnamefont{Parameswaran}},
  \bibinfo{author}{\bibfnamefont{A.~M.} \bibnamefont{Turner}},
  \bibinfo{author}{\bibfnamefont{D.~P.} \bibnamefont{Arovas}},
  \bibnamefont{and}
  \bibinfo{author}{\bibfnamefont{A.}~\bibnamefont{Vishwanath}},
  \bibinfo{journal}{Nature Physics} \textbf{\bibinfo{volume}{9}},
  \bibinfo{pages}{299} (\bibinfo{year}{2013}).

\bibitem[{\citenamefont{Steinberg et~al.}(2014)\citenamefont{Steinberg, Young,
  Zaheer, Kane, Mele, and Rappe}}]{Steinberg}
\bibinfo{author}{\bibfnamefont{J.~A.} \bibnamefont{Steinberg}},
  \bibinfo{author}{\bibfnamefont{S.~M.} \bibnamefont{Young}},
  \bibinfo{author}{\bibfnamefont{S.}~\bibnamefont{Zaheer}},
  \bibinfo{author}{\bibfnamefont{C.~L.} \bibnamefont{Kane}},
  \bibinfo{author}{\bibfnamefont{E.~J.} \bibnamefont{Mele}}, \bibnamefont{and}
  \bibinfo{author}{\bibfnamefont{A.~M.} \bibnamefont{Rappe}},
  \bibinfo{journal}{Phys. Rev. Lett.} \textbf{\bibinfo{volume}{112}},
  \bibinfo{pages}{036403} (\bibinfo{year}{2014}).

\bibitem[{\citenamefont{Yang and Nagaosa}(2014)}]{YangNagaosa}
\bibinfo{author}{\bibfnamefont{B.-J.} \bibnamefont{Yang}} \bibnamefont{and}
  \bibinfo{author}{\bibfnamefont{N.}~\bibnamefont{Nagaosa}},
  \bibinfo{journal}{Nature Communications} \textbf{\bibinfo{volume}{5}},
  \bibinfo{pages}{4898} (\bibinfo{year}{2014}).

\bibitem[{\citenamefont{Young and Kane}(2015)}]{YoungKane}
\bibinfo{author}{\bibfnamefont{S.~M.} \bibnamefont{Young}} \bibnamefont{and}
  \bibinfo{author}{\bibfnamefont{C.~L.} \bibnamefont{Kane}},
  \bibinfo{journal}{Phys. Rev. Lett.} \textbf{\bibinfo{volume}{115}},
  \bibinfo{pages}{126803} (\bibinfo{year}{2015}).

\bibitem[{\citenamefont{Fang et~al.}(2015)\citenamefont{Fang, Chen, Kee, and
  Fu}}]{Fang2015}
\bibinfo{author}{\bibfnamefont{C.}~\bibnamefont{Fang}},
  \bibinfo{author}{\bibfnamefont{Y.}~\bibnamefont{Chen}},
  \bibinfo{author}{\bibfnamefont{H.-Y.} \bibnamefont{Kee}}, \bibnamefont{and}
  \bibinfo{author}{\bibfnamefont{L.}~\bibnamefont{Fu}}, \bibinfo{journal}{Phys.
  Rev. B} \textbf{\bibinfo{volume}{92}}, \bibinfo{pages}{081201}
  (\bibinfo{year}{2015}).

\bibitem[{\citenamefont{Watanabe et~al.}(2016)\citenamefont{Watanabe, Po,
  Zaletel, and Vishwanath}}]{Watanabe2016}
\bibinfo{author}{\bibfnamefont{H.}~\bibnamefont{Watanabe}},
  \bibinfo{author}{\bibfnamefont{H.~C.} \bibnamefont{Po}},
  \bibinfo{author}{\bibfnamefont{M.~P.} \bibnamefont{Zaletel}},
  \bibnamefont{and}
  \bibinfo{author}{\bibfnamefont{A.}~\bibnamefont{Vishwanath}},
  \bibinfo{journal}{Phys. Rev. Lett.} \textbf{\bibinfo{volume}{117}},
  \bibinfo{pages}{096404} (\bibinfo{year}{2016}).

\bibitem[{\citenamefont{Bradlyn et~al.}(2016)\citenamefont{Bradlyn, Cano, Wang,
  Vergniory, Felser, Cava, and Bernevig}}]{Bradlyn2016}
\bibinfo{author}{\bibfnamefont{B.}~\bibnamefont{Bradlyn}},
  \bibinfo{author}{\bibfnamefont{J.}~\bibnamefont{Cano}},
  \bibinfo{author}{\bibfnamefont{Z.}~\bibnamefont{Wang}},
  \bibinfo{author}{\bibfnamefont{M.~G.} \bibnamefont{Vergniory}},
  \bibinfo{author}{\bibfnamefont{C.}~\bibnamefont{Felser}},
  \bibinfo{author}{\bibfnamefont{R.~J.} \bibnamefont{Cava}}, \bibnamefont{and}
  \bibinfo{author}{\bibfnamefont{B.~A.} \bibnamefont{Bernevig}},
  \bibinfo{journal}{Science} \textbf{\bibinfo{volume}{353}}
  (\bibinfo{year}{2016}).

\bibitem[{\citenamefont{Chen et~al.}(2016)\citenamefont{Chen, Kim, and
  Kee}}]{Chen2016}
\bibinfo{author}{\bibfnamefont{Y.}~\bibnamefont{Chen}},
  \bibinfo{author}{\bibfnamefont{H.-S.} \bibnamefont{Kim}}, \bibnamefont{and}
  \bibinfo{author}{\bibfnamefont{H.-Y.} \bibnamefont{Kee}},
  \bibinfo{journal}{Phys. Rev. B} \textbf{\bibinfo{volume}{93}},
  \bibinfo{pages}{155140} (\bibinfo{year}{2016}).

\bibitem[{\citenamefont{Wieder and Kane}(2016)}]{Wieder}
\bibinfo{author}{\bibfnamefont{B.~J.} \bibnamefont{Wieder}} \bibnamefont{and}
  \bibinfo{author}{\bibfnamefont{C.~L.} \bibnamefont{Kane}},
  \bibinfo{journal}{Phys. Rev. B} \textbf{\bibinfo{volume}{94}},
  \bibinfo{pages}{155108} (\bibinfo{year}{2016}).

\bibitem[{\citenamefont{Fang et~al.}(2016)\citenamefont{Fang, Weng, Dai, and
  Fang}}]{ChenFangReview}
\bibinfo{author}{\bibfnamefont{C.}~\bibnamefont{Fang}},
  \bibinfo{author}{\bibfnamefont{H.}~\bibnamefont{Weng}},
  \bibinfo{author}{\bibfnamefont{X.}~\bibnamefont{Dai}}, \bibnamefont{and}
  \bibinfo{author}{\bibfnamefont{Z.}~\bibnamefont{Fang}},
  \bibinfo{journal}{Chinese Physics B} \textbf{\bibinfo{volume}{25}},
  \bibinfo{pages}{117106} (\bibinfo{year}{2016}).

\bibitem[{\citenamefont{Yang et~al.}(2017)\citenamefont{Yang, Bojesen,
  Morimoto, and Furusaki}}]{YangBojesen}
\bibinfo{author}{\bibfnamefont{B.-J.} \bibnamefont{Yang}},
  \bibinfo{author}{\bibfnamefont{T.~A.} \bibnamefont{Bojesen}},
  \bibinfo{author}{\bibfnamefont{T.}~\bibnamefont{Morimoto}}, \bibnamefont{and}
  \bibinfo{author}{\bibfnamefont{A.}~\bibnamefont{Furusaki}},
  \bibinfo{journal}{Phys. Rev. B} \textbf{\bibinfo{volume}{95}},
  \bibinfo{pages}{075135} (\bibinfo{year}{2017}).

\bibitem[{\citenamefont{{Chang} et~al.}(2016)\citenamefont{{Chang}, {Sanchez},
  {Wieder}, {Xu}, {Schindler}, {Belopolski}, {Huang}, {Singh}, {Wu}, {Neupert}
  et~al.}}]{Titus2016}
\bibinfo{author}{\bibfnamefont{G.}~\bibnamefont{{Chang}}},
  \bibinfo{author}{\bibfnamefont{D.~S.} \bibnamefont{{Sanchez}}},
  \bibinfo{author}{\bibfnamefont{B.~J.} \bibnamefont{{Wieder}}},
  \bibinfo{author}{\bibfnamefont{S.-Y.} \bibnamefont{{Xu}}},
  \bibinfo{author}{\bibfnamefont{F.}~\bibnamefont{{Schindler}}},
  \bibinfo{author}{\bibfnamefont{I.}~\bibnamefont{{Belopolski}}},
  \bibinfo{author}{\bibfnamefont{S.-M.} \bibnamefont{{Huang}}},
  \bibinfo{author}{\bibfnamefont{B.}~\bibnamefont{{Singh}}},
  \bibinfo{author}{\bibfnamefont{D.}~\bibnamefont{{Wu}}},
  \bibinfo{author}{\bibfnamefont{T.}~\bibnamefont{{Neupert}}},
  \bibnamefont{et~al.}, \bibinfo{journal}{ArXiv e-prints}
  (\bibinfo{year}{2016}), \eprint{1611.07925}.

\bibitem[{\citenamefont{{Wang} et~al.}(2017)\citenamefont{{Wang}, {Jian}, and
  {Yao}}}]{Wang2017}
\bibinfo{author}{\bibfnamefont{L.}~\bibnamefont{{Wang}}},
  \bibinfo{author}{\bibfnamefont{S.-K.} \bibnamefont{{Jian}}},
  \bibnamefont{and} \bibinfo{author}{\bibfnamefont{H.}~\bibnamefont{{Yao}}},
  \bibinfo{journal}{ArXiv e-prints}  (\bibinfo{year}{2017}),
  \eprint{1702.06140}.

\bibitem[{\citenamefont{Kakihana et~al.}(2015)\citenamefont{Kakihana, Teruya,
  Nishimura, Nakamura, Takeuchi, Haga, Harima, Hedo, Nakama, and
  Onuki}}]{Kakihana2015}
\bibinfo{author}{\bibfnamefont{M.}~\bibnamefont{Kakihana}},
  \bibinfo{author}{\bibfnamefont{A.}~\bibnamefont{Teruya}},
  \bibinfo{author}{\bibfnamefont{K.}~\bibnamefont{Nishimura}},
  \bibinfo{author}{\bibfnamefont{A.}~\bibnamefont{Nakamura}},
  \bibinfo{author}{\bibfnamefont{T.}~\bibnamefont{Takeuchi}},
  \bibinfo{author}{\bibfnamefont{Y.}~\bibnamefont{Haga}},
  \bibinfo{author}{\bibfnamefont{H.}~\bibnamefont{Harima}},
  \bibinfo{author}{\bibfnamefont{M.}~\bibnamefont{Hedo}},
  \bibinfo{author}{\bibfnamefont{T.}~\bibnamefont{Nakama}}, \bibnamefont{and}
  \bibinfo{author}{\bibfnamefont{Y.}~\bibnamefont{Onuki}},
  \bibinfo{journal}{Journal of the Physical Society of Japan}
  \textbf{\bibinfo{volume}{84}}, \bibinfo{pages}{094711}
  (\bibinfo{year}{2015}).

\bibitem[{\citenamefont{Teruya et~al.}(2016)\citenamefont{Teruya, Suzuki, Aoki,
  Honda, Nakamura, Nakashima, Amako, Harima, Hedo, Nakama et~al.}}]{Teruya2016}
\bibinfo{author}{\bibfnamefont{A.}~\bibnamefont{Teruya}},
  \bibinfo{author}{\bibfnamefont{F.}~\bibnamefont{Suzuki}},
  \bibinfo{author}{\bibfnamefont{D.}~\bibnamefont{Aoki}},
  \bibinfo{author}{\bibfnamefont{F.}~\bibnamefont{Honda}},
  \bibinfo{author}{\bibfnamefont{A.}~\bibnamefont{Nakamura}},
  \bibinfo{author}{\bibfnamefont{M.}~\bibnamefont{Nakashima}},
  \bibinfo{author}{\bibfnamefont{Y.}~\bibnamefont{Amako}},
  \bibinfo{author}{\bibfnamefont{H.}~\bibnamefont{Harima}},
  \bibinfo{author}{\bibfnamefont{M.}~\bibnamefont{Hedo}},
  \bibinfo{author}{\bibfnamefont{T.}~\bibnamefont{Nakama}},
  \bibnamefont{et~al.}, \bibinfo{journal}{Journal of the Physical Society of
  Japan} \textbf{\bibinfo{volume}{85}}, \bibinfo{pages}{064716}
  (\bibinfo{year}{2016}).

\bibitem[{\citenamefont{Bouhon and Black-Schaffer}(2017)}]{Bouhon}
\bibinfo{author}{\bibfnamefont{A.}~\bibnamefont{Bouhon}} \bibnamefont{and}
  \bibinfo{author}{\bibfnamefont{A.~M.} \bibnamefont{Black-Schaffer}},
  \bibinfo{journal}{Phys. Rev. B} \textbf{\bibinfo{volume}{95}},
  \bibinfo{pages}{241101} (\bibinfo{year}{2017}).

\bibitem[{\citenamefont{Bzdu\v{s}ek et~al.}(2016)\citenamefont{Bzdu\v{s}ek, Wu,
  R\"{u}egg, Sigrist, and Soluyanov}}]{Bzdusek2016}
\bibinfo{author}{\bibfnamefont{T.}~\bibnamefont{Bzdu\v{s}ek}},
  \bibinfo{author}{\bibfnamefont{Q.}~\bibnamefont{Wu}},
  \bibinfo{author}{\bibfnamefont{A.}~\bibnamefont{R\"{u}egg}},
  \bibinfo{author}{\bibfnamefont{M.}~\bibnamefont{Sigrist}}, \bibnamefont{and}
  \bibinfo{author}{\bibfnamefont{A.~A.} \bibnamefont{Soluyanov}},
  \bibinfo{journal}{Nature} \textbf{\bibinfo{volume}{538}}, \bibinfo{pages}{75}
  (\bibinfo{year}{2016}).

\bibitem[{\citenamefont{Yang et~al.}(2015)\citenamefont{Yang, Morimoto, and
  Furusaki}}]{BJYang}
\bibinfo{author}{\bibfnamefont{B.-J.} \bibnamefont{Yang}},
  \bibinfo{author}{\bibfnamefont{T.}~\bibnamefont{Morimoto}}, \bibnamefont{and}
  \bibinfo{author}{\bibfnamefont{A.}~\bibnamefont{Furusaki}},
  \bibinfo{journal}{Phys. Rev. B} \textbf{\bibinfo{volume}{92}},
  \bibinfo{pages}{165120} (\bibinfo{year}{2015}).

\bibitem[{\citenamefont{Son and Spivak}(2013)}]{Son}
\bibinfo{author}{\bibfnamefont{D.~T.} \bibnamefont{Son}} \bibnamefont{and}
  \bibinfo{author}{\bibfnamefont{B.~Z.} \bibnamefont{Spivak}},
  \bibinfo{journal}{Phys. Rev. B} \textbf{\bibinfo{volume}{88}},
  \bibinfo{pages}{104412} (\bibinfo{year}{2013}).

\bibitem[{\citenamefont{Chan et~al.}(2016)\citenamefont{Chan, Chiu, Chou, and
  Schnyder}}]{Chan2016}
\bibinfo{author}{\bibfnamefont{Y.-H.} \bibnamefont{Chan}},
  \bibinfo{author}{\bibfnamefont{C.-K.} \bibnamefont{Chiu}},
  \bibinfo{author}{\bibfnamefont{M.~Y.} \bibnamefont{Chou}}, \bibnamefont{and}
  \bibinfo{author}{\bibfnamefont{A.~P.} \bibnamefont{Schnyder}},
  \bibinfo{journal}{Phys. Rev. B} \textbf{\bibinfo{volume}{93}},
  \bibinfo{pages}{205132} (\bibinfo{year}{2016}).

\end{thebibliography}

\end{document}